\documentclass[aps,prd,twocolumn,superscriptaddress,nofootinbib]{revtex4-2}

\usepackage{amsmath,amssymb}
\usepackage{graphicx}
\usepackage{bm}
\usepackage[colorlinks=true,linkcolor=blue,citecolor=blue,urlcolor=blue]{hyperref}

\begin{document}

\title{Modified Entanglement Patterns in Two-Flavor Neutrinos from Quantum-Gravity Interactions}

\author{Bipin Singh Koranga}
\email{bskoranga@kmc.du.ac.in}
\author{Pranav Kumar}
\author{Baktiar Wasir Farooq}
\affiliation{Department of Physics, Kirori Mal College, University of Delhi, Delhi-110007, India}

\date{\today}

\begin{abstract}
In this work, we investigate the influence of quantum-gravity--induced corrections on the entanglement entropy associated with two-flavor neutrino oscillations in vacuum. Using the von Neumann entropy as a measure of quantum correlations, we analyze how Planck-scale--suppressed modifications --- implemented through quantum-gravity--motivated changes in the neutrino dispersion relation --- affect the evolution of entanglement during successive oscillation cycles. By performing a statistical study over a range of oscillation lengths and neutrino energies, we identify characteristic deviations in the entropy profile arising from quantum-gravity effects. Our results suggest that entanglement entropy provides a sensitive probe of small quantum-gravity--induced departures from standard neutrino oscillation dynamics, highlighting its potential role in testing Planck-scale physics within neutrino phenomenology.
\end{abstract}

\maketitle

\section{Introduction}

Quantum entanglement in neutrino oscillations has emerged as an important topic in recent years, offering a novel perspective on the quantum structure of flavor mixing and propagation \cite{blasone2008a,blasone2008b,blasone2013,blasone2014,capolupo2021}. Earlier studies have extensively examined entanglement entropy as a tool to characterize quantum correlations between neutrino mass and flavor modes. Blasone and collaborators pioneered much of this development by formulating mode entanglement in terms of flavor transition probabilities and exploring its implications within both quantum mechanical and quantum field--theoretic frameworks \cite{martin2016,siwach2021,gautam2020,cervia2020,mallick2022}. Subsequent works expanded these ideas to multiparty systems, three-flavor oscillations, and dense astrophysical environments, demonstrating strong connections between entanglement measures, flavor coherence, and experimentally accessible observables \cite{amelino1998,hossenfelder2013,lambiase2005,anchordoqui2005,ellis2000}. In parallel, quantum-gravity (QG) phenomenology has provided several mechanisms through which Planck-scale effects can modify neutrino propagation. These include modified dispersion relations, generalized uncertainty principles (GUP), Lorentz-violating terms, and stochastic space-time fluctuations, all of which can introduce small but potentially observable deviations in oscillation probabilities \cite{minkowski1977,yanagida1979,weinberg1979,vissani2003}. Given that entanglement entropy is highly sensitive to subtle changes in the phase structure and coherence properties of neutrino states, it provides a promising tool for probing such QG-induced corrections.

In this work, we study the entanglement entropy between neutrino mass eigenstates in vacuum oscillations, incorporating quantum-gravity--motivated modifications to the oscillation parameters. Using the von Neumann entropy as the primary entanglement measure, we analyze how Planck-scale corrections influence the evolution of entanglement over successive oscillation cycles and across different values of the ratio $L/E$. Our findings highlight how even small QG-induced shifts in the oscillation phase can lead to detectable deviations in the entanglement profile.

The structure of this paper is as follows. In Sec.~\ref{sec:mixing}, we review the standard two-flavor oscillation framework and introduce the quantum-gravity modification employed in our analysis. Section~\ref{sec:entropy} presents the calculation of the von Neumann entanglement entropy and its evolution as a function of $L/E$. Section~\ref{sec:results} discusses the numerical results and the implications of entanglement observables as probes of Planck-scale physics. Finally, Sec.~\ref{sec:conclusions} provides concluding remarks.

\section{Neutrino Mixing Parameters from Quantum-Gravity Effects}
\label{sec:mixing}

In this section, we highlight that the neutrino mass-squared differences may receive contributions from physics operating above the Grand Unified Theory (GUT) scale, particularly through gravitational interactions involving neutrinos. While the dominant structure of the neutrino mass matrix is generally attributed to GUT-scale dynamics --- most commonly realized through the seesaw mechanism \cite{minkowski1977,yanagida1979} --- Planck-suppressed gravitational effects can introduce additional corrections. These corrections manifest as small but significant modifications to the neutrino mass-squared differences. The effective gravitational interaction between neutrinos and the Higgs field can be formulated as an $SU(2)_L \times U(1)_Y$ gauge-invariant dimension-5 operator \cite{weinberg1979}:
\begin{equation}
\mathcal{L}_{\rm grav} = \frac{\lambda_{\alpha\beta}}{M_{\rm pl}} (\psi_{A\alpha}\epsilon\psi_C)C^{-1}_{ab}(\psi_{B\beta}\epsilon_{BD}\psi_D) + {\rm h.c.}
\label{eq:lgrav}
\end{equation}
Here and throughout we use Greek indices $\alpha,\beta$ for the flavor states and Latin indices $i,j,k$ for the mass states. In the above equation $\psi_\alpha = (\nu_\alpha, l_\alpha)$ is the lepton doublet, $\phi = (\phi^+,\phi^0)$ is the Higgs doublet, and $M_{\rm pl} = 1.2\times 10^{19}\,{\rm GeV}$ is the Planck mass. $\lambda$ is a $3\times 3$ matrix in flavor space with each element $\mathcal{O}(1)$. The Lorentz indices $a,b=1,2,3,4$ are contracted with the charge-conjugation matrix $C$, and the $SU(2)_L$ isospin indices $A,B,C,D=1,2$ are contracted with $\epsilon = i\sigma_2$, where $\sigma_m$ ($m=1,2,3$) are the Pauli matrices. After spontaneous electroweak symmetry breaking, the Lagrangian in Eq.~(\ref{eq:lgrav}) generates an additional term in the neutrino mass matrix,
\begin{equation}
\mathcal{L}_{\rm mass} = \frac{v^2}{M_{\rm pl}}\lambda_{\alpha\beta}\,\nu_\alpha C^{-1}\nu_\beta,
\label{eq:lmass}
\end{equation}
where $v = 174\,{\rm GeV}$ is the vacuum expectation value of electroweak symmetry breaking. We assume that the gravitational interaction is ``flavor blind,'' i.e., $\lambda_{\alpha\beta}$ is independent of the $\alpha,\beta$ indices. Thus the Planck-scale contribution to the neutrino mass matrix is
\begin{equation}
\mu\lambda = \mu
\begin{pmatrix}
1 & 1 & 1 \\
1 & 1 & 1 \\
1 & 1 & 1
\end{pmatrix},
\label{eq:mulambda}
\end{equation}
where the scale $\mu$ is
\begin{equation}
\mu = \frac{v^2}{M_{\rm pl}} = 2.5\times 10^{-6}\,{\rm eV}.
\label{eq:mu}
\end{equation}
We take Eq.~(\ref{eq:mulambda}) as a perturbation to the main part of the neutrino mass matrix generated by GUT dynamics. We treat $M$ as the unperturbed (zeroth-order) mass matrix in the mass eigenbasis. Let $U$ be the mixing matrix at zeroth order. The corresponding zeroth-order mass matrix in flavor space \cite{vissani2003} is given by
\begin{equation}
M = U^* \,{\rm diag}(M_i)\,U^\dagger,
\label{eq:M0}
\end{equation}
where $U_{\alpha i}$ is the usual mixing matrix and $M_i$ are the neutrino masses generated by the Grand Unified Theory. Most of the parameters related to neutrino oscillation are known; the major uncertainty resides in the mixing element $U_{e3}$. We adopt the usual parametrization
\begin{align}
\frac{|U_{e2}|}{|U_{e1}|} &= \tan\theta_{12}, \label{eq:t12}\\
\frac{|U_{\mu 3}|}{|U_{\tau 3}|} &= \tan\theta_{23}, \label{eq:t23}\\
|U_{e3}| &= \sin\theta_{13}. \label{eq:t13}
\end{align}
In terms of the above mixing angles, the mixing matrix is
\begin{equation}
\begin{split}
U = {\rm diag}(e^{if_1},e^{if_2},e^{if_3})\,R(\theta_{23})\,\Delta\,R(\theta_{13})\\
{}\times\Delta^*\,R(\theta_{12})\,{\rm diag}(e^{ia_1},e^{ia_2},1).
\end{split}
\label{eq:Umatrix}
\end{equation}
The matrix $\Delta = {\rm diag}(e^{i\delta/2},1,e^{-i\delta/2})$ contains the Dirac phase, which leads to CP violation in neutrino oscillations. $a_1$ and $a_2$ are the so-called Majorana phases, which affect neutrinoless double-beta decay. The phases $f_1,f_2,f_3$ are usually absorbed into the definition of the charged-lepton field. Due to Planck-scale effects on neutrino mixing, the new mixing matrix can be written as \cite{koranga2008}
\begin{equation}
U' = U(1+i\delta\theta),
\label{eq:Uprime}
\end{equation}
with
\begin{equation}
U =
\begin{pmatrix}
U_{e1} & U_{e2} & U_{e3} \\
U_{\mu 1} & U_{\mu 2} & U_{\mu 3} \\
U_{\tau 1} & U_{\tau 2} & U_{\tau 3}
\end{pmatrix},
\end{equation}
and the perturbation contributes explicitly as
\begin{equation}
\begin{split}
i\Big[&U_{e2}\delta\theta_{12}^{*}+U_{e3}\delta\theta_{23}^{*},\ U_{e1}\delta\theta_{12}+U_{e3}\delta\theta_{23}^{*},\\
&U_{e1}\delta\theta_{13}+U_{e3}\delta\theta_{23}^{*};\ldots\Big],
\end{split}
\label{eq:deltaU}
\end{equation}
where the full matrix structure follows the same pattern for the $\mu$ and $\tau$ rows. Here $U' = U+\delta U$ denotes the effective neutrino mixing matrix including Planck-scale corrections, where $U$ is the standard PMNS mixing matrix and $\delta U$ is the perturbative modification induced by quantum-gravity effects, and $\delta\theta$ is a Hermitian matrix that is first order in $\mu$ \cite{koranga2008}. The first-order mass-squared difference $\Delta M_{ij}^2 = M_i^2 - M_j^2$ gets modified \cite{koranga2010} as
\begin{equation}
\Delta'_{ij} = \Delta_{ij} + 2\big(M_i\,{\rm Re}(m_{ii}) - M_j\,{\rm Re}(m_{jj})\big),
\label{eq:Deltaprime}
\end{equation}
where
\begin{equation}
m = \mu\, U^\dagger \lambda\, U, \qquad \mu = \frac{v^2}{M_{\rm pl}} = 2.5\times 10^{-6}\,{\rm eV}.
\end{equation}
The change in the elements of the mixing matrix, parametrized by $\delta\theta$ \cite{vissani2003}, is given by
\begin{equation}
\delta\theta_{ij} = \frac{i\,{\rm Re}(m_{jj})(M_i+M_j) - {\rm Im}(m_{jj})(M_i-M_j)}{\Delta M_{ij}'^2}.
\label{eq:deltathetaij}
\end{equation}
Equation~(\ref{eq:deltathetaij}) determines only the off-diagonal elements of the matrix $\delta\theta_{ij}$; the diagonal elements can be set to zero by phase invariance.

Using Eq.~(\ref{eq:Uprime}), we can calculate the neutrino mixing angles arising from Planck-scale effects,
\begin{align}
\frac{|U'_{e2}|}{|U'_{e1}|} &= \tan\theta'_{12}, \label{eq:t12p}\\
\frac{|U'_{\mu 3}|}{|U'_{\tau 3}|} &= \tan\theta'_{23}, \label{eq:t23p}\\
|U'_{e3}| &= \sin\theta'_{13}. \label{eq:t13p}
\end{align}

For nearly degenerate neutrinos, $M_3-M_1 \cong M_3-M_2 \gg M_2-M_1$, since $\Delta_{31}\cong \Delta_{32}\gg\Delta_{21}$. In terms of the modified mixing matrix elements, this simplifies \cite{koranga2008} to
\begin{equation}
\begin{split}
\tan\theta'_{12} = \frac{|U'_{e2}|}{|U'_{e1}|} = \tan\theta_{12}\\
{}+ \frac{2\mu M\,|z_1|^2|z_2|^2}{\Delta M_{21}^2\cos^2\theta_{12}}\cos(a_1+a_2)\cos(a_1-a_2).
\end{split}
\label{eq:tanprime}
\end{equation}
The modified mass-squared difference simplifies \cite{koranga2011} to
\begin{equation}
\begin{split}
\Delta'_{21} &= \Delta_{21} + 2\big(M_1{\rm Re}(m_{11}) - M_2{\rm Re}(m_{22})\big)\\
&= \Delta_{21} + 2\mu M\big[|z_2|^2\cos(2a_1) - |z_1|^2\cos(2a_2)\big].
\end{split}
\label{eq:Deltaprime21}
\end{equation}
We define the complex number $z_i = U_{ei}+U_{\mu i}+U_{\tau i}$, where $U_{\alpha i}$ is in general a function of all six phases.

\section{Entanglement Entropy for Neutrino Oscillations due to Quantum Gravity}
\label{sec:entropy}

The von Neumann entropy is a fundamental tool for quantifying entanglement in bipartite quantum systems. In the present work, the quantum-gravity contribution is implemented within an effective field theory (EFT) framework. We consider Planck-scale-suppressed higher-dimensional operators that modify neutrino propagation through small perturbations to the neutrino mass matrix and dispersion relation. Such corrections are equivalent to leading-order Planck-suppressed operators arising from quantum-gravity-motivated scenarios and can phenomenologically mimic Lorentz-invariance-violating or deformed-dispersion effects at low energies. The corrections are treated perturbatively, ensuring that standard neutrino oscillation physics is recovered in the limit $M_{\rm pl}\to\infty$. Therefore, the analysis does not rely on a specific microscopic model of quantum gravity but rather on a generic EFT parametrization capturing the leading observable consequences of Planck-scale physics.

For a density matrix $\rho$, the von Neumann entropy is defined as \cite{blasone2009}
\begin{equation}
S(\rho) = -{\rm Tr}(\rho\log\rho),
\label{eq:vonNeumann}
\end{equation}
where ${\rm Tr}\,\rho=1$ for a normalized density matrix, and $\rho$ is the density matrix of the neutrino flavor state,
\begin{equation}
\rho = |\nu_\alpha(t)\rangle\langle\nu_\alpha(t)|.
\label{eq:rho}
\end{equation}
We restrict the discussion to two neutrino flavors, denoted $\nu_\alpha$ and $\nu_\beta$, which form an $SU(2)$ isospin doublet in flavor space \cite{sarkar2001}. Under the assumption that flavor occupation behaves as a reference quantum number, these states can be mapped to two-qubit basis vectors,
\begin{equation}
|\nu_\alpha\rangle = |1,0\rangle = \begin{pmatrix}0\\1\\0\\0\end{pmatrix}, \qquad
|\nu_\beta\rangle = |0,1\rangle = \begin{pmatrix}0\\0\\1\\0\end{pmatrix}.
\label{eq:qubits}
\end{equation}
Even for a single particle, flavor superposition naturally leads to entanglement between flavor modes.

\subsection{Von Neumann Entropy from Oscillation Probabilities}

We consider only two-flavor neutrino oscillations. The two mass eigenstates $\nu_1$ and $\nu_2$ are linear combinations of the flavor states $\nu_e$ and $\nu_\mu$,
\begin{align}
\nu_e &= \cos\theta\,\nu_1 + \sin\theta\,\nu_2, \\
\nu_\mu &= -\sin\theta\,\nu_1 + \cos\theta\,\nu_2.
\end{align}
In the two-flavor scheme, the $\nu_e\to\nu_\mu$ oscillation probability is given by
\begin{equation}
P_{e\mu}({\rm osc.}) = \sin^2 2\theta_{12}\,\sin^2\!\left(1.27\,\frac{\Delta_{21}L}{E}\right),
\label{eq:Posc}
\end{equation}
and the $\nu_e\to\nu_e$ survival probability is
\begin{equation}
P_{ee}({\rm surv.}) = 1-\sin^2 2\theta_{12}\,\sin^2\!\left(1.27\,\frac{\Delta_{21}L}{E}\right),
\label{eq:Psurv}
\end{equation}
where $\Delta_{21}=m_2^2-m_1^2$ is in ${\rm eV}^2$, the baseline length $L$ is in km, and the neutrino energy $E$ is in GeV.

For a two-flavor system, the reduced density matrix of a flavor mode leads to an effective two-level statistical mixture, and the resulting entanglement entropy is
\begin{equation}
S(\rho) = -P_{\rm surv}\log P_{\rm surv} - P_{\rm osc}\log P_{\rm osc},
\label{eq:Srho}
\end{equation}
where $P_{\rm surv}$ is the probability that the neutrino remains in its initial flavor and $P_{\rm osc}$ is the probability that it converts to the other flavor. Thus the entanglement entropy directly reflects the neutrino flavor dynamics.

\subsection{Quantum-Gravity-Affected Entanglement Entropy}

In the presence of quantum-gravity (QG) corrections, the effective mixing angle and mass-squared splitting receive modifications, resulting in QG-corrected oscillation probabilities $P_{\rm surv}^{\rm QG}$ and $P_{\rm osc}^{\rm QG}$. The entanglement entropy incorporating quantum-gravity effects becomes
\begin{equation}
S^{\rm QG}(\rho) = -P_{\rm surv}^{\rm QG}\log P_{\rm surv}^{\rm QG} - P_{\rm osc}^{\rm QG}\log P_{\rm osc}^{\rm QG}.
\label{eq:SQG}
\end{equation}
In this work, the effective parameters are modified by the quantum-gravity deviations discussed above; these influence the oscillation behavior in an analogous way by altering the effective mixing angle, changing the oscillation length, and shifting the mass-squared differences.

\section{Numerical Results}
\label{sec:results}

The correction term to the mass-squared difference depends crucially on the structure of the neutrino mass spectrum. For both the normal and inverted hierarchical spectra, the Planck-scale-induced correction is extremely small and can be safely neglected. We therefore adopt a degenerate neutrino mass spectrum and take the common neutrino mass to be $m_\nu = 2\,{\rm eV}$, consistent with the upper bound obtained from tritium beta-decay experiments \cite{kraus2005}. Incorporating Planck-scale effects, the modified mass-squared difference and effective mixing angle take the form \cite{king2004,sarkar2001}
\begin{align}
\Delta'_{21} &= \Delta_{21} + 2\mu M\big[|z_2|^2\cos(2a_1)-|z_1|^2\cos(2a_2)\big], \label{eq:num1}\\
\tan\theta'_{12} &= \tan\theta_{12} + \frac{2\mu M|z_1|^2|z_2|^2}{\Delta M_{21}^2\cos^2\theta_{12}}\notag\\
&\qquad{}\times\cos(a_1+a_2)\cos(a_1-a_2). \label{eq:num2}
\end{align}
These expressions illustrate how Planck-scale operators induce small but theoretically significant corrections to the neutrino oscillation parameters, particularly under the assumption of a nearly degenerate mass spectrum.

The contribution of the Planck-scale correction term, $\epsilon = 2\big(M_i{\rm Re}(m_{11})-M_j{\rm Re}(m_{22})\big)$, can be either additive or subtractive depending sensitively on the values of the Majorana phases $a_1,a_2$ and the charged-lepton phases $f_i$. In our numerical analysis, we employ the standard two-neutrino mixing parameters. The experimental value of the mass-squared difference is taken as $\Delta_{21}=8.0\times10^{-5}\,{\rm eV}^2$ \cite{ahmad2001}. For simplicity, we set the charged-lepton phases to $f_1=f_2=f_3=0$; with $\theta_{13}=0$, the Dirac CP-violating phase $\delta$ does not contribute to the zeroth-order leptonic mixing matrix. We take $\theta_{12}=34^\circ$ \cite{fukuda1998}.

\begin{figure}[t]
\centering
\includegraphics[width=\columnwidth]{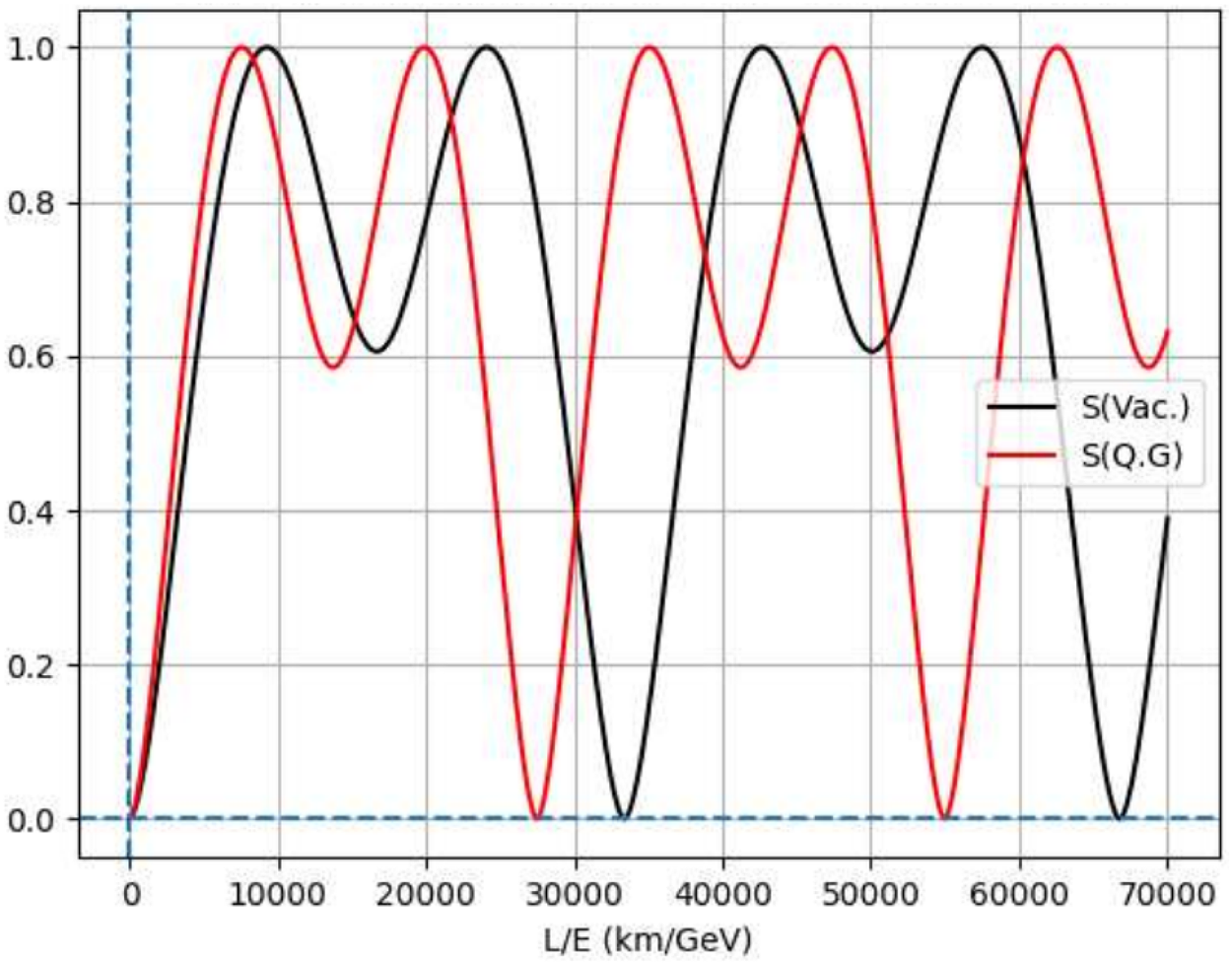}
\caption{Entropy comparison in vacuum and under quantum-gravity corrections, for $\theta'_{12}=34^\circ$ in the quantum-gravity region.}
\label{fig:fig1}
\end{figure}

\begin{figure}[t]
\centering
\includegraphics[width=\columnwidth]{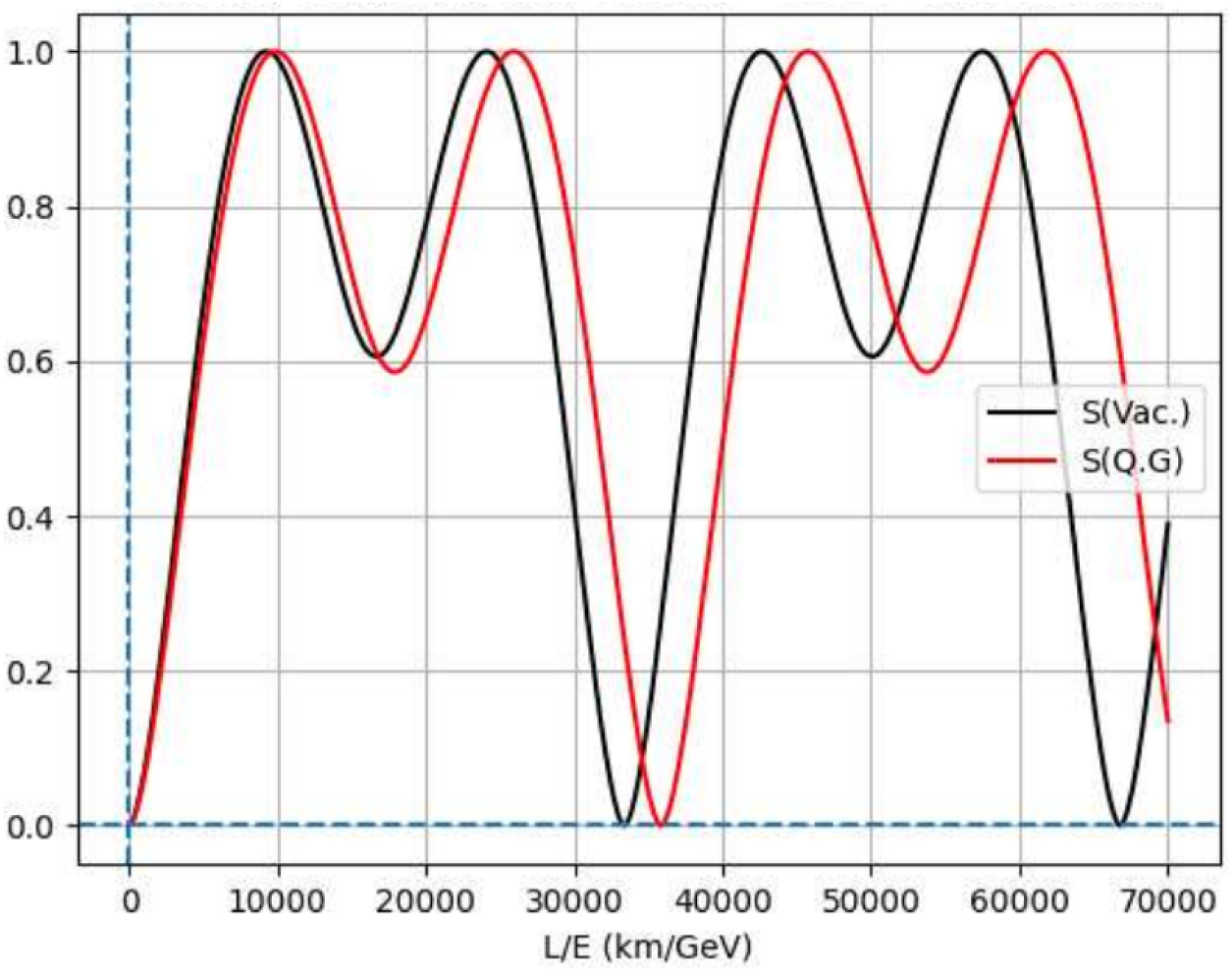}
\caption{Entropy comparison in vacuum and under quantum-gravity corrections, for $\theta'_{12}=33.99^\circ$ in the quantum-gravity region.}
\label{fig:fig2}
\end{figure}

Based on Eq.~(\ref{eq:tanprime}), we examine the entanglement entropy for neutrino oscillations in the presence of quantum gravity. According to the formalism described above, flavor neutrino states can be conceptualized as entangled superpositions of the mass qubits $|\nu_i\rangle$ at any given moment, where the entanglement depends only on the mixing angle and the neutrino mass-squared difference. The entanglement entropy for two-flavor neutrino oscillation in vacuum, together with the quantum-gravity-corrected entropy, is displayed in Figs.~\ref{fig:fig1} and~\ref{fig:fig2}.

\textit{Choice of numerical parameters.} The numerical values used in the figures correspond to representative points chosen within experimentally allowed parameter ranges. The mixing angles $33.99^\circ$ and $34^\circ$ were selected because they lie within the currently allowed interval for the mixing angle and are close to maximal mixing. Using two nearby values allows us to illustrate the sensitivity of the entanglement entropy to small variations of the oscillation parameters in the experimentally preferred region; the qualitative behavior of the results remains unchanged across the full allowed parameter space.

\textit{Experimental constraints.} The parameters describing the Planck-scale corrections are not taken arbitrarily; rather, they are restricted to ranges compatible with existing experimental and observational bounds from solar, atmospheric, reactor, and long-baseline neutrino oscillation experiments, as well as astrophysical neutrino observations. Current phenomenological studies constrain such corrections to extremely small magnitudes, and the parameters adopted in our numerical analysis lie well within these allowed limits. The results should therefore be interpreted as physical benchmark scenarios rather than as unconstrained parameter choices.

It was observed that the quantum-gravity entanglement-entropy area under the $L/E$ curve is smaller than that of the vacuum case, suggesting that the degree of entanglement is more efficient under quantum-gravity corrections. Greater entanglement corresponds to more coherent oscillation at the Planck scale; the more effective coherent-oscillation region, $L/E \le \pi/(1.27\,\Delta_{21})$, was identified directly from the graphs.

\subsection{Comparison Between Vacuum and Quantum-Gravity Entropy}

As discussed in Ref.~\cite{koranga2008} regarding the matter-induced modification of entropy slopes for two-flavor neutrino oscillations in matter and vacuum, we performed a similar comparison here. In Fig.~\ref{fig:fig1} we compare the von Neumann entropy for the vacuum case with that obtained including quantum-gravity effects (for $\theta'_{12}=34^\circ$). In this plot, the peaks of the entanglement entropy under quantum gravity converge relative to the vacuum case. Since $L\approx t$ in natural units ($c=1$), an increase in the effective mass-squared difference $\Delta m^2$ enhances the phase-accumulation rate $\phi = (\Delta m^2/4E)\,L$. Consequently, the oscillation probability reaches $P=1/2$ at a smaller propagation length $L$, leading to an earlier attainment of the entropy maximum. This indicates that the maximally mixed state of $|\nu_e\rangle$ is reached earlier than in vacuum, so the slope of the entanglement entropy over the same range of $L/E$ is larger in the quantum-gravity case, confirming that the neutrino oscillations proceed more rapidly under quantum gravity. For $\theta'_{12}=34^\circ$, the mass-squared difference modifies to $\Delta'_{21}=9\times 10^{-5}\,{\rm eV}^2$ \cite{koranga2008}, whereas for the vacuum case with $\theta_{12}=33.4^\circ$ the mass-squared difference is $\Delta_{21}=7.6\times 10^{-5}\,{\rm eV}^2$. This shift in the modified mass-squared difference produces the convergence of the entropy curves observed in the quantum-gravity case.

In Fig.~\ref{fig:fig2} we performed a similar comparison but with $\theta'_{12}=33.99^\circ$. Here, the von Neumann entropy curve diverges relative to the vacuum case. The same reasoning applies: for $\theta'_{12}=33.99^\circ$ the mass-squared difference modifies to $\Delta'_{21}=6.9\times 10^{-5}\,{\rm eV}^2$ \cite{koranga2008}, whereas for the vacuum case with $\theta_{12}=33.4^\circ$ the mass-squared difference remains $\Delta_{21}=7.6\times 10^{-5}\,{\rm eV}^2$. Because this modified mass-squared difference is smaller than the vacuum value, the oscillation probabilities in the quantum-gravity case are reduced relative to the vacuum case, producing the observed divergence.

\section{Conclusions}
\label{sec:conclusions}

We have studied the influence of quantum gravity on two-flavor neutrino oscillation probabilities and the associated entanglement entropy. As illustrated in Figs.~\ref{fig:fig1} and~\ref{fig:fig2}, the entanglement entropy undergoes a noticeable shift when Planck-scale-induced corrections are incorporated into the neutrino mass-squared differences and mixing angles. These quantum-gravity effects enhance the variation in entanglement entropy across the oscillation cycle, indicating that Planck-scale physics modifies the degree of mode entanglement in the two-flavor neutrino system. A change in entanglement entropy over a given interval of $L/E$ reflects the continuous redistribution of quantum correlations between flavor modes, corresponding to the persistence of flavor oscillations; in the presence of quantum-gravity corrections, this behavior becomes more pronounced. The entropy of the system is computed using the von Neumann entropy, which quantifies how the modified oscillation probabilities --- arising from Planck-scale perturbations to the neutrino parameters --- alter the entanglement structure of the neutrino state. Thus, quantum gravity not only perturbs the neutrino oscillation parameters but also imprints a measurable signature on the entanglement entropy, offering a potential probe of Planck-scale physics through neutrino oscillations.

\section*{CRediT authorship contribution statement}
\textbf{Bipin Singh Koranga:} Writing -- review \& editing, Writing -- original draft, Supervision, Conceptualization. \textbf{Baktiar Wasir Farooq:} Writing -- review \& editing, Writing -- original draft, Conceptualization. \textbf{Pranav Kumar:} Resources.

\section*{Declaration of Interest}
The authors declare that they have no known competing financial interests or personal relationships that could have appeared to influence the work reported in this paper, with the following exception: Baktiar Wasir Farooq reports that administrative support, article publishing charges, equipment, statistical analysis, travel, and writing assistance were provided by the University of Delhi, Kirori Mal College, and reports a non-financial relationship with the same institution. All other authors declare no known competing financial interests or personal relationships that could have appeared to influence the work reported in this paper.

\section*{Data Availability Statement}
No data were associated with this manuscript.

\end{document}